\documentclass[
twocolumn,
]{ceurart}

\usepackage[T1]{fontenc}
\usepackage[utf8]{inputenc}
\usepackage[english, norsk]{babel}
    
\begin{document}

\selectlanguage{english}

\copyrightyear{2021}
\copyrightclause{Copyright for this paper by its authors.
  Use permitted under Creative Commons License Attribution 4.0
  International (CC BY 4.0).}

\conference{Joint Proceedings of the ACM IUI 2021 Workshops,
  April 13--17, 2021, College Station, USA}

\title{Changing Salty Food Preferences with Visual and Textual Explanations in a Search Interface}

\author[1]{Arngeir Berge}[%
]

\ead{arngeir.berge@norceresearch.no}
\address[1]{NORCE Norwegian Research Centre,
  P.O.Box 22 Nygårdstangen, 5838 Bergen, Norway}

\author[2]{Vegard Velle Sjoen}[
]
\address[2]{University of Bergen, P.O.Box 7800, 5020 Bergen, Norway}
\ead{vsj010@uib.no}

\author[2,3]{Alain Starke}[
]
\ead{Alain.Starke@uib.no}

\address[3]{Wageningen University \& Research, 6708 PB Wageningen, The Netherlands}


\author[2]{Christoph Trattner}[
]
\ead{Christoph.Trattner@uib.no}

\begin{abstract}
Salt is consumed at too high levels in the general population, causing high blood pressure and related health problems. In this paper, we present results of ongoing research that tries to reduce salt intake via technology and in particular from an interface perspective. In detail, this paper features results of a study that examines the extent to which visual and textual explanations in a search interface can change salty food preferences. An online user study with 200 participants demonstrates that this is possible in food search results by accompanying recipes with a visual taste map that includes salt-replacer herbs and spices in the calculation of salty taste.
\end{abstract}


\maketitle

\section{Introduction}

Worldwide obesity levels are increasing~\cite{Trattner2017a}. A main issue is the high prevalence of unhealthy foods available, both offline and online~\cite{starke2019recsys,Trattner2017}. On the other hand, technology has been shown to be useful to tackle obesity. Food recommender technology, for example, has demonstrated the potential to change people's eating behavior~\cite{Trattner2017a, musto2020towards}. Yet, less explored is the principle of ``search'', the main means of finding information about food on the Web.

To contribute to this little researched area, we have focused on the extent to which search interfaces can change people's eating preferences. One recent study shows healthy recipes can be boosted by presenting attractive food images alongside them \cite{starke2021}, overcoming possible innate preferences for unhealthy food \cite{Trattner2017}.


In this work, we focus on salty food preferences and how to change these during the search process. The main principle is to replace food that has a high salt content with less salty food. The food items under investigation are online food recipes. To support possible changes in user preferences, we emphasize the preference of salt replacers in food through visual and textual explanations. Most people who like salty taste in food are equally satisfied if salt is reduced slightly, as long as substitute herbs or spices are added~\cite{Durack2008}. Such salt replacers (SRs) are ingredients like oregano and garlic, which mimic the taste sensation triggered by consuming salt. To promote salt-replacer recipes (SR recipes) we test an approach in which the visual and textual explanations of food are altered in a search interface. The intention of our explanations is to boost SR recipes by educating the user about the healthier nutritional content. Such an intervention is referred to as a ``Boost'', a type of nudging that promotes small changes in behavior through education~\cite{Hertwig2017}.

\textbf{Main objective:} We seek to show that visual and textual explanations of salt content and salt replacers (SRs) in a search interface can alter salty food preferences. Focusing on the online recipe domain, we posit the following research questions:
\begin{itemize}
\item \textit{RQ1:} To what extent do salty food preferences change due to visual and textual explanations on a recipe's salty taste?
\item \textit{RQ2:} To what extent do other recipe and user characteristics affect salty food preferences?
\end{itemize}

In the remainder of this paper, we first discuss relevant work related to this research (cf. Section~\ref{sec:background}), positioning our approach. Then, we describe the contents of our user study, including recipe dataset and our visual and textual explanations for salty taste. Finally, we describe our main results and discuss their implications.

\section{Background}\label{sec:background}
In this section, we first introduce key health implications of consuming food high in salt content. Secondly, we present studies that have examined how food preferences can be shifted towards healthier options (i.e. digital nudging), by focusing on \textit{how} food is presented rather than changing \textit{what} is presented. Thirdly, we show how this work is based on earlier research in the field, and where it makes a contribution.

\subsection{Salt in Food and Salt Replacers}
High salt/sodium content in food is one of the fundamental indirect causes of disease and death~\cite{he2009comprehensive, yoon_sodium_2013}. Cardiovascular diseases can be attributed to increased blood pressure, which is found to be responsible for 62\% of strokes and 49\% of coronary hearth diseases. Starting a downward spiral, a person's blood pressure is increased by high salt intake~\cite{he2009comprehensive}, which can often be attributed to high levels of consumption of preprocessed food~\cite{Trattner2017a}.

The high prevalence of salt in food is also notable online. Trattner et al.~\cite{Trattner2017a} show that most recipes on the world's largest recipe website, Allrecipes.com, are relatively unhealthy. To make matters worse, in terms of health, most personalization algorithms tend to prioritize popular, yet unhealthy items in their search results, which tend to surpass WHO and FSA guidelines for salt intake.

The healthiness of online recipes can be improved by swapping salt content for replacement ingredients that mimic the taste. Salt-replacer (SR) herbs and spices can do this and have health benefits in their own right, sometimes even lowering blood pressure~\cite{Taladrid2020}. Examples of such ingredients are garlic, oregano and rosemary, which can be found in various online recipes~\cite{Trattner2017a}. However, personalization algorithms in search have yet to prioritize recipes that contain SR ingredients (SR recipes) so that users can reap their health benefits.

\subsection{Digital Food Nudges}
Besides changing \textit{what} recipes are recommended, one way to promote SR recipes is to highlight them in a choice interface. To change the choice context without altering what is presented can be referred to as (digital) nudging. Cadario et al.~\cite{Cadario2020} classify three types of nudges: cognitively-oriented (e.g. educating the user about nutritional values), affectively-oriented (e.g. affecting a user's feelings by putting happy faces on healthy products), and behaviorally-oriented nudges (e.g. changing how options are presented in a supermarket shelf). The latter is found to be the most effective offline. This suggests that healthy alternatives can be promoted in a digital interface, such as a search user interface (UI), by re-ranking recipes based on health. In addition, accompanying recipes with a cognitively-oriented explanation about which ingredients can replace salt content might also ``nudge'' users towards healthy options, as the health benefits of SR ingredients are relatively unknown~\cite{pietrasik2015effect}. This might affect healthy choices in the long term.

A specific type of nudging is called ``Boosting''. ``Boosts'' promote small changes in behavior, by educating users about the suggested changes at the same time~\cite{Hertwig2017}. We believe that it is possible to both make it easier for users to select healthy recipes and increase their knowledge, by calculating a wholesome salty taste score based on research about SRs, as well as educating users about how selecting SRs can accomplish a salty taste with less sodium.

\subsection{Differences to Previous Research}

The key differences between this research and the works mentioned above are the focus on the type of interface, the food preferences, and the type of nudges being used to change food preferences.

With regard to user interface design, a lot of research is devoted to recommender interfaces~\cite{Trattner2017}, while our research is on food search. Another key difference is the type of preferences we consider. While previous research examined food preferences in general~\cite{starke2019recsys}, we focus on salty food preferences, as high salt intake is shown to be the main cause for current cardiovascular diseases~\cite{he2009comprehensive}. Finally, compared to previous work, this study is among the first to investigate the use of taste maps and textual explanations to change people's salty food preferences.


\section{Methods}\label{sec:methods}

This section presents the methods used and applied in our research. In particular, we discuss our recipe dataset sample, reveal details about the visual and textual explanations and how we developed these and, finally, describe the design of our online study and how this study was carried out.

\subsection{Recipe Dataset}

To address our research questions, we selected main courses from a recipe database, obtained from the recipe website Allrecipes.com and provided by~\cite{Trattner2017a}. The dataset comprised a sub-sample of 1031 recipes with detailed information about ingredients, directions and nutritional values. To perform our research we selected two times six recipes corresponding to two search result sets in our prototype, see Table~\ref{tab:selected-recipes}. Six of the recipes answered a search for ``chicken'' and the other six answered a search for ``pork''. Each of the search result sets had half-half recipes with and without SR ingredients. When selecting SR recipes we looked for SRs that appear in the literature~\cite{Durack2008, Ghawi2014, Mitchell2013, Taladrid2020} and ended up with recipes containing one or more of the SR ingredients garlic, rosemary or oregano. The search result sets were composed to reflect the variety found in real-world food sites by covering a wide spectrum of sodium content measured per serving. Each search result set also mirrored the large dataset in that some recipes contained SR ingredients and some did not. In our case, three SR recipes and three Non-SR recipes. 

Table~\ref{tab:selected-recipes} shows the recipes for the two search result sets used in the online user study. For SR recipes we calculated an SR-boosting score. For Non-SR recipes we let the SR-boosting score be similar to the sodium score.

\begin{table*}
  \caption{The selected recipes in the chicken and pork search result sets with sodium score, salt-replacer-boosting score and salt-replacer ingredients. Bold numbers show where the sodium score has been boosted. Note: We have marked recipes without salt replacers like for example garlic with N/A.}
  \label{tab:selected-recipes}
      \begin{tabular}{l|lccr}
        \toprule
        Recipe sets & Recipe title & Sodium score & SR-boosting score & Salt replacers\\
        \midrule
        \multirow{6}{*}{Chicken} & Slow Cooker Sweet and Tangy Chicken & 0.81 & \textbf{0.97} & Garlic\\
        & Tomato Chicken Parmesan & 0.69 & 0.69 & N/A\\
        & Chicken Scarpariello & 0.61 & \textbf{0.73} & Rosemary, garlic\\
        & Oh-So-Good Chicken & 0.4 & 0.4 & N/A\\
        & Chicken Marsala & 0.3 & \textbf{0.36} & Oregano\\
        & Baked Lemon Chicken with Mushroom Sauce & 0.11 & 0.11 & N/A\\
      \cline{1-1}
      \multirow{6}{*}{Pork} & North Carolina-Style Pulled Pork & 0.94 & 0.94 & N/A\\
        & Pork Chops for the Slow Cooker & 0.74 & \textbf{0.89} & Garlic\\
        & Skillet Pork Chops with Potatoes and Onion & 0.6 & 0.6 & N/A\\
        & Slow Cooker Teriyaki Pork Tenderloin & 0.41 & \textbf{0.49} & Garlic\\
        & Roast Pork with Maple and Mustard Glaze & 0.29 & 0.29 & N/A\\
        & Pork Marsala & 0.2 & \textbf{0.24} & Oregano, garlic\\
      \bottomrule
      \end{tabular}
\end{table*}


\subsection{Designing Salty Taste Explanations}

\begin{figure}[]
  \centering
  \vspace{0.9cm}
  \includegraphics[width=0.8\linewidth]{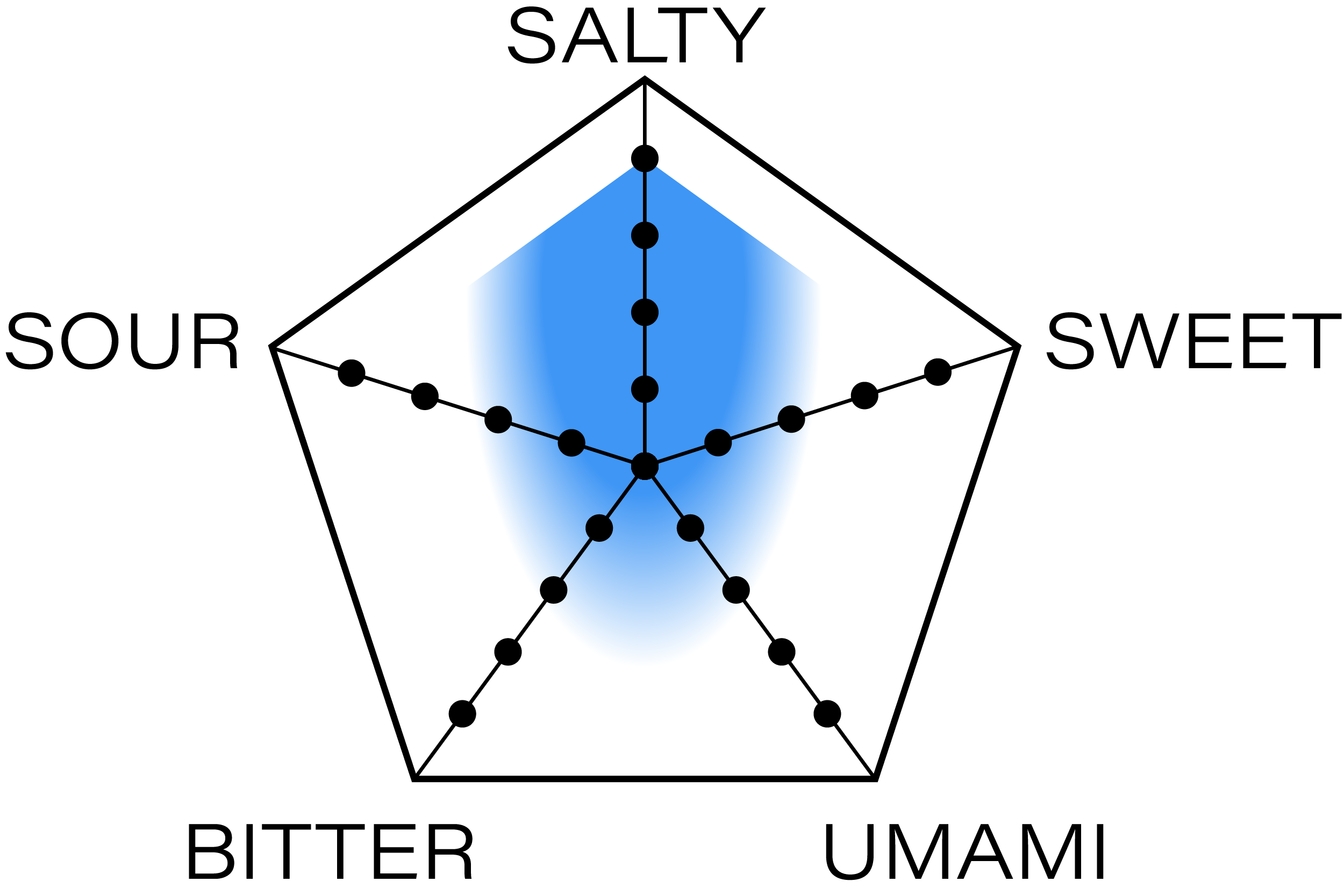}
  \vspace{0.3cm}
  \caption{Example of a taste map presented for each recipe. Only the salty taste axis is calculated in this study and therefore the coloration fades towards the other axes.}
  \label{figure:taste-map}
  \vspace{0.3cm}
\end{figure}

We developed a visual explanation of taste that we call a taste map, see Figure~\ref{figure:taste-map}. The taste map was inspired by other visual taste descriptions in the literature~\cite{FryVennerod2018}. Our visual taste explanation is a pentagram with one axis for each of the five basic tastes. For this study, we only calculated values for the salty taste axis. Had values for all axes been calculated, a shape would emerge to show a complete taste profile that would enable users to look for a variety of visual shapes when browsing search results or recipe presentations.


The taste map covered a scale from 0 to 1 on the salty axis, with a granularity of 0.1. We anticipated that many of our participants would expect a recipe with a mean sodium level to be displayed with 0.5, the middle value of our scale. The sodium content per serving in the larger dataset was normally distributed. The mean was 0.9 grams and the 95 percentile was 2.0 grams. We therefore made it simple by regarding the maximum sodium level as the mean times two, which is 1.8 grams. Any serving with 1.8 grams sodium or more would turn up as 1 on our scale. A mean sodium level would turn up as 0.5 on our scale. In one of the experiment conditions, the value of the taste map was boosted when shown alongside SR recipes. The taste map functioned as a visual salty taste explanation for all recipes at all times in our experiment. In the SR-boosting condition, however, the taste map had a heightened explanatory role by showing visually that SRs contribute to the perceived salty taste. Calculating the contribution of salt replacers to salty taste is a subject of food chemistry and beyond the scope of this paper. To be able to perform our experiment, we took a pragmatic approach and added 20\% to the sodium score to calculate an SR-boosting score for SR recipes.

In addition to the visual explanation, we designed a textual explanation in our search prototype to boost SR recipes, as depicted on the bottom right of Figure~\ref{figure:conditions}. The textual explanation had two parts: First, the headline ``Healthy ingredients enhancing salty taste'', followed by an explanation like ``Oregano and garlic joins the salt in enriching the flavors of this recipe.''

\subsection{Online User Study}
We used our recipe dataset and salty taste explanations in our online user study. 

\subsubsection{Participants}

To have a diverse pool of participants, we used two crowdsourcing platforms. We recruited a total of 200 participants: 100 on Prolific, 100 on Amazon MTurk. We only sampled U.S. nationals, as ingredient amounts were denoted in U.S. metrics. As is common in research, we had a different required approval rate for the two platforms~\cite{schild_behavior_2020}, 90\% and 98\% respectively. Prolific participants were reimbursed with 2 GBP for participation as we assumed that our study took 14 minutes, while ``MTurkers'' were compensated with 0.5 USD when we found out that the mean completion time was much shorter (7 minutes). The eventual sample comprised 54.8\% males, with a mean age of 37.6 years ($SD=14.16$). 

\subsubsection{Prototype}

\begin{figure}[]
  \centering
  \includegraphics[width=1.0\linewidth]{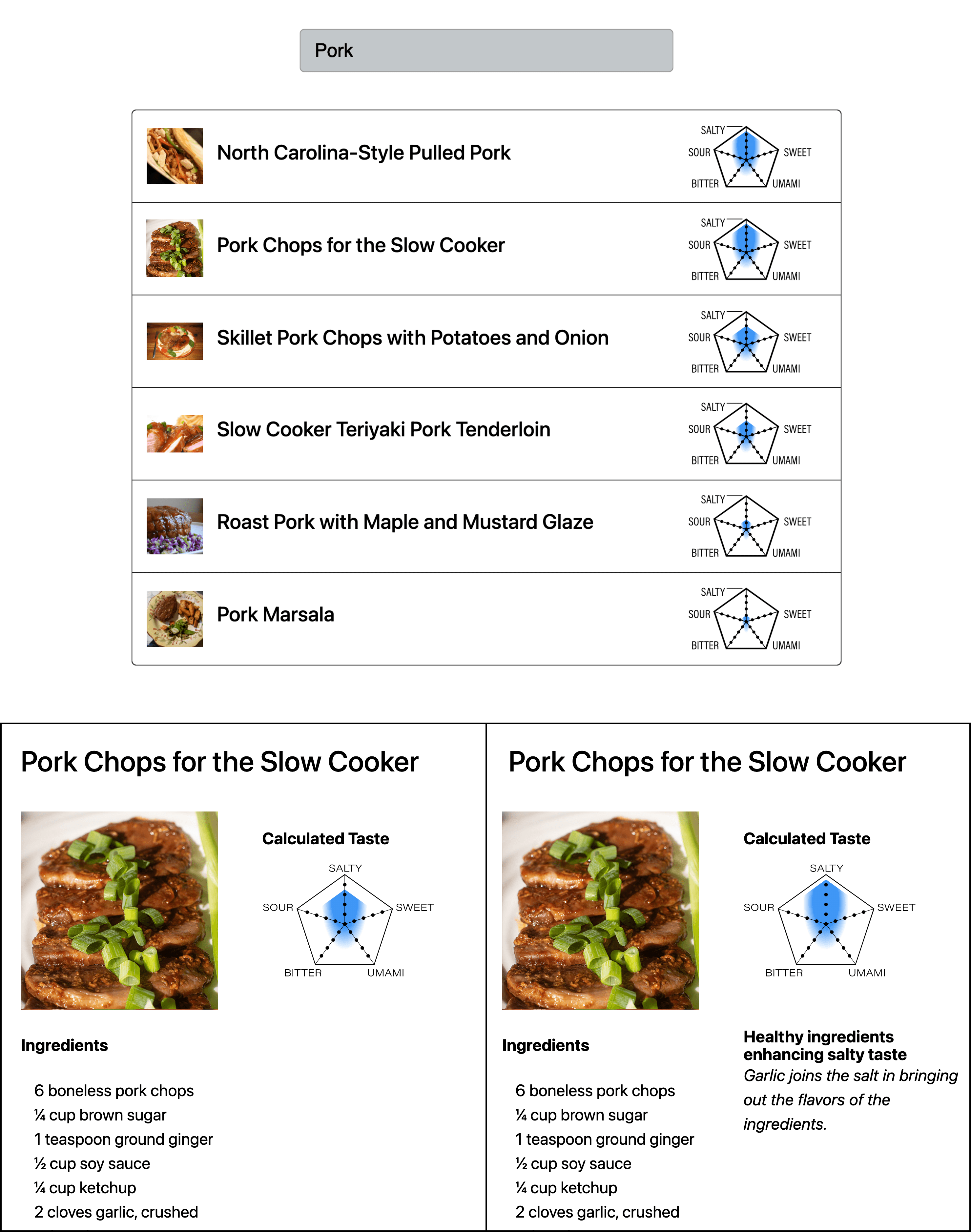}
  \caption{Top: One of the two search result sets. Bottom left and right: A single participant would see this recipe in only one of the left/right presentation states. Since the recipe had salt-replacer ingredients, it could be presented in two states: On the left with a taste map based on sodium score and on the right with a textual explanation and a taste map based on SR-boosting score. Non-SR recipes could only have the state shown on the left-hand side.}
  \label{figure:conditions}
\end{figure}

The prototype had web pages for consent and questions before and after the experiment. More notably, it had web pages for search samples (see upper part of Figure~\ref{figure:conditions}). A search bar prefilled with a search term was shown together with six search results. To the right of each recipe title was a miniature taste map, so that users could quickly scan the taste of each recipe in the results. The prototype also had separate web pages to present the full recipes that appeared in the search results (see lower part of Figure~\ref{figure:conditions}). Each page contained the recipe title, an image of the meal, ingredients, cooking directions and nutritional info, in addition to a taste map section. Figure~\ref{figure:conditions} also shows side-by-side the two ways SR recipes were presented. On the left-hand side the salty taste value in the taste map was based on sodium score. On the right-hand side, the salty taste value in the taste map was boosted visually and accompanied by a textual explanation.

\subsubsection{Research Design}

\begin{figure}[]
  \centering
  \includegraphics[width=1\linewidth]{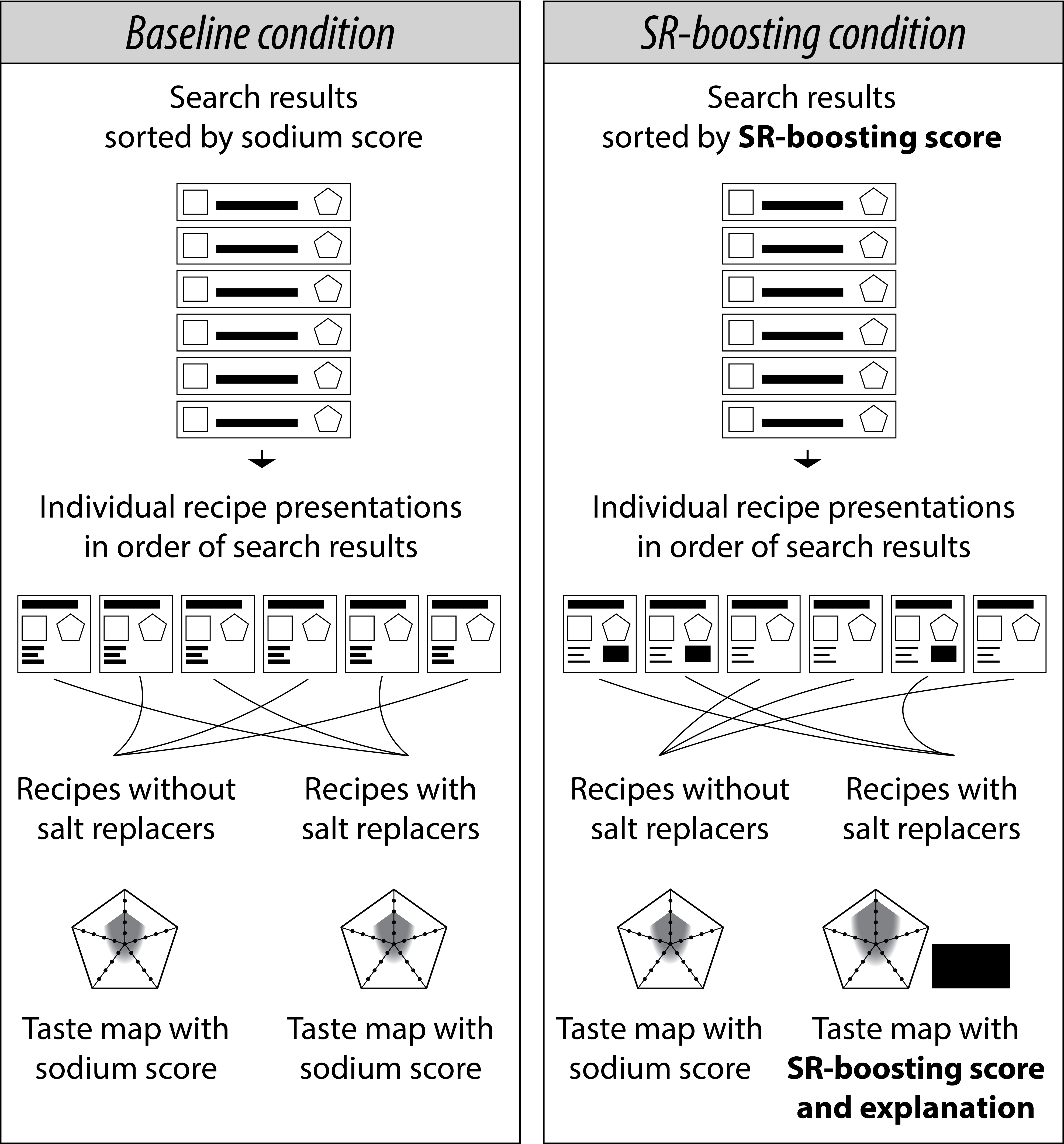}
  \caption{The two main conditions. SR recipes and Non-SR recipes were present in both conditions, but were treated differently. Non-SR recipes were never boosted, but SR recipes were boosted in the SR-boosting condition.}
  \label{figure:sorting-and-boosting}
\end{figure}

\textbf{Recipe search results and individual presentations.} In the two following main conditions, users were first shown a sample search with a list of six search results containing a mixture of three Non-SR recipes, and three SR recipes. The search results were ordered from high to low value on the miniature taste maps. Subsequently, the participants were guided through recipe presentation pages for each search result in the list. On top of each recipe, the participants were asked: ``Please rate this recipe according to how it fulfills your salty food preference on a scale from 1 to 7 (1 means very poorly and 7 means very well).'' Participants could glean salty taste levels from the taste map, the ingredients, as well as the nutritional info section's sodium value and percentage of daily recommended intake.

\textbf{Baseline Condition.} This was one of two main conditions in the experiment. In this condition a search for either ``chicken'' or ``pork'' was shown, the opposite term from that in the other main condition. Search results were ranked according to the recipes' sodium score (as opposed to SR-boosting score), displayed on the miniature taste maps. The participants were then guided through the recipe presentations. Three of the six recipes contained SR ingredients, but participants were not given any textual explanation about this, and there was no boosting of the taste map value in search results or on the recipe pages.

\textbf{SR-Boosting Condition.} This was the other main condition. A search for either ``chicken'' or ``pork'' was presented, the opposite term from that in the baseline condition. Search results were ranked according to the recipes' SR-boosting score as shown in Table~\ref{tab:selected-recipes}. Three of the search results contained SR ingredients and were boosted. This would most often make a visible difference in the miniature taste map beside the search result and in the larger taste map in the recipe presentation. In the presentation of SR recipes the taste map was accompanied by a textual explanation about the presence of SR ingredients. For example the headline: ``Healthy ingredients enhancing salty taste'', and text: ``Garlic joins the salt in bringing out the flavor of the ingredients''. The SR-boosting condition only boosted SR recipes. Non-SR recipes did not change, had no boosted taste map value and no textual explanation. Although the presentation of Non-SR recipes did not change across conditions, they were present in the search result sets so that each set mirrored the mixture of recipe types typically found in a real world search results.

\subsubsection{Procedure}

\begin{figure}[]
  \centering
  \includegraphics[width=1\linewidth]{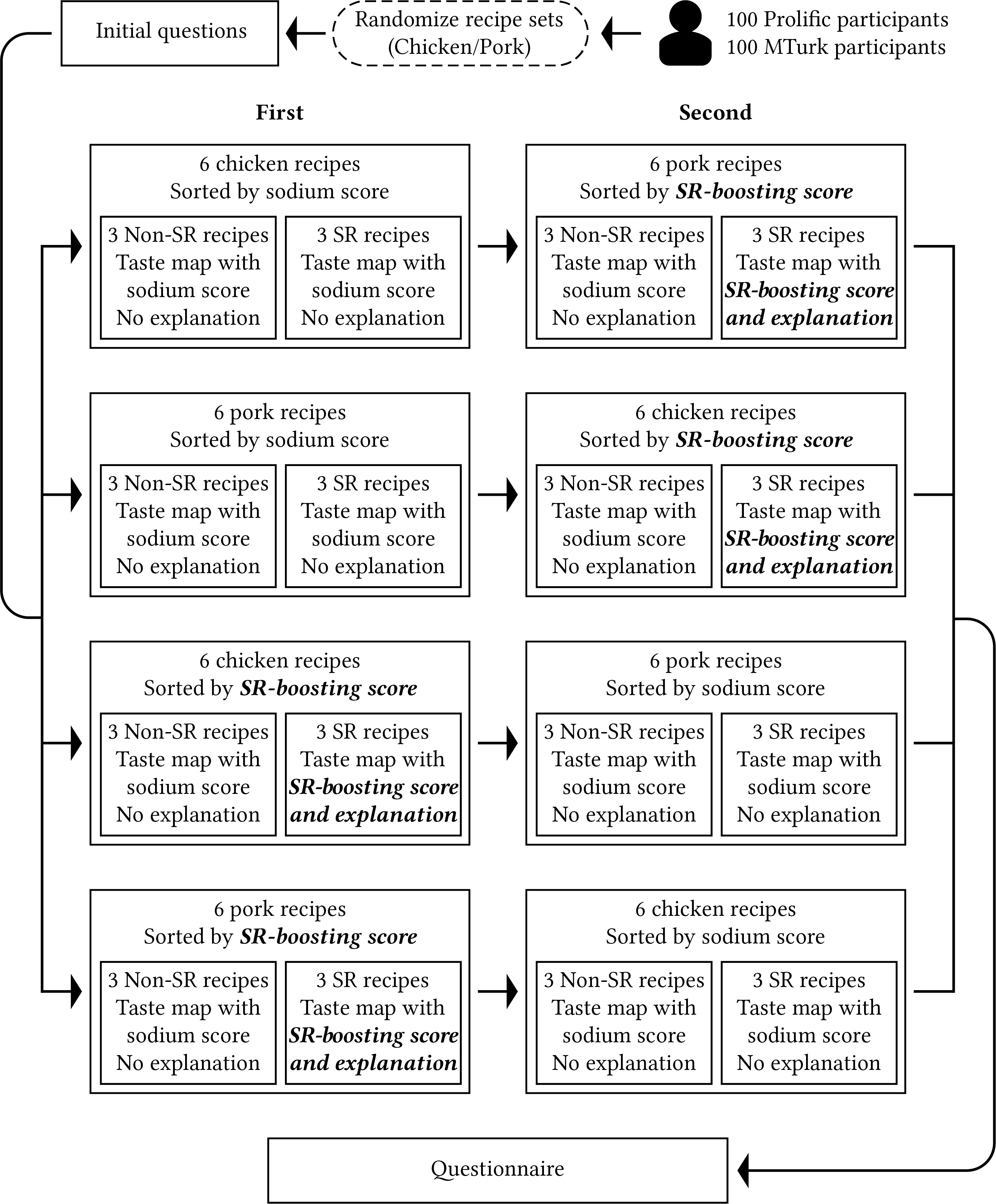}
  \caption{Full procedure of the online study including the within-subjects research design. In the sequence of two searches, the search result lists could be sorted by either sodium score or SR-boosting score and corresponding miniature taste maps were shown. In individual recipe presentations following each search result list, presentations of SR recipes were sometimes boosted with an explanation and a taste map value based on SR-boosting score. Boosting is marked with italics.}
  \label{figure:study-flow}
\end{figure}

Figure~\ref{figure:study-flow} shows that after initial questions the participants were randomly assigned to four branches, ending up with a questionnaire. There were two main conditions with randomized order. These were paired with the two search samples, which also had randomized order. The visual scores in the taste maps were based on sodium scores in the baseline condition and SR-boosting scores in the SR-boosting condition. The latter condition also showed a textual explanation accompanying a boosted taste map value when a recipe had SR ingredients.

Initially in the procedure, participants were asked about characteristics including cooking experience and healthy eating habits. They were then shown a static page with an example of how a search in our recipe site could look. Either the search for ``chicken'' or ``pork'' was shown together with the corresponding recipe set consisting of six search results. Participants were asked to go through each recipe, read the recipe presentation, and rate it according to how well it satisfied their salty food preference. A second search sample showed the remaining search term, either ``chicken'' or ``pork'', and the resulting recipe set. Once more, participants were asked to go through the recipes of the search results and rate them. Finally, participants filled out a short questionnaire where they were asked about their attitude towards SRs.

\subsubsection{Measures}

The answers to the questions participants were asked before the experiment were used as independent variables later in the analysis. We inquired about their age and gender (Male, Female, Other). Then, about what their highest completed education was: Less than high school, High school or equivalent, Bachelor degree (e.g. BA, BSc), Master degree, Doctorate, or Prefer not to say. They were also asked about their cooking experience and eating habits on a 5-point scale from ``Very low / Very unhealthy'' to ``Very high / Very healthy''.

In the experiment we collected user preference data used as dependent variable in the analysis. Each participant was shown a total of 12 full recipe presentations. For each recipe the user was asked to read the presentation and rate the recipe \textit{according to how it fulfilled their salty-food preference} on a 7-point Likert scale from ``Very poorly'' to ``Very well''.

The questionnaire after the experiment notably asked a control variable Likert scale question: ``Research shows that many people can be just as satisfied if salt is reduced a little and garlic, oregano or rosemary added. To what extent do you think that you would be satisfied with this in the future on a scale from 1 (to no extent) to 7 (to a high extent)?''


\section{Results}\label{sec:results}

In this section we outline the main results from the online study, regarding our two research questions.

\subsection{RQ1: Food Preferences \& Salt Replacement Explanations}


We first examined whether user preferences for SR recipes changed due to our taste map value boost and textual explanation (RQ1). Overall, 200 participants gave 2x 600 ratings for Non-SR recipes and 2x 600 ratings for SR recipes across the two main conditions. Since the boosting-explanations were only presented alongside recipes that actually contained SR ingredients, we performed a dependent t-test that compared the 2x 600 ratings for SR recipes. The t-test showed that ratings given in the SR-boosting condition ($M=4.54$, $SD=1.77$) were significantly higher than in the baseline condition ($M=4.33$, $SD=1.79$): $t$(1198) = 2.08, $p<0.05$. 

\begin{figure}[]
  \centering
  \includegraphics[width=1\linewidth]{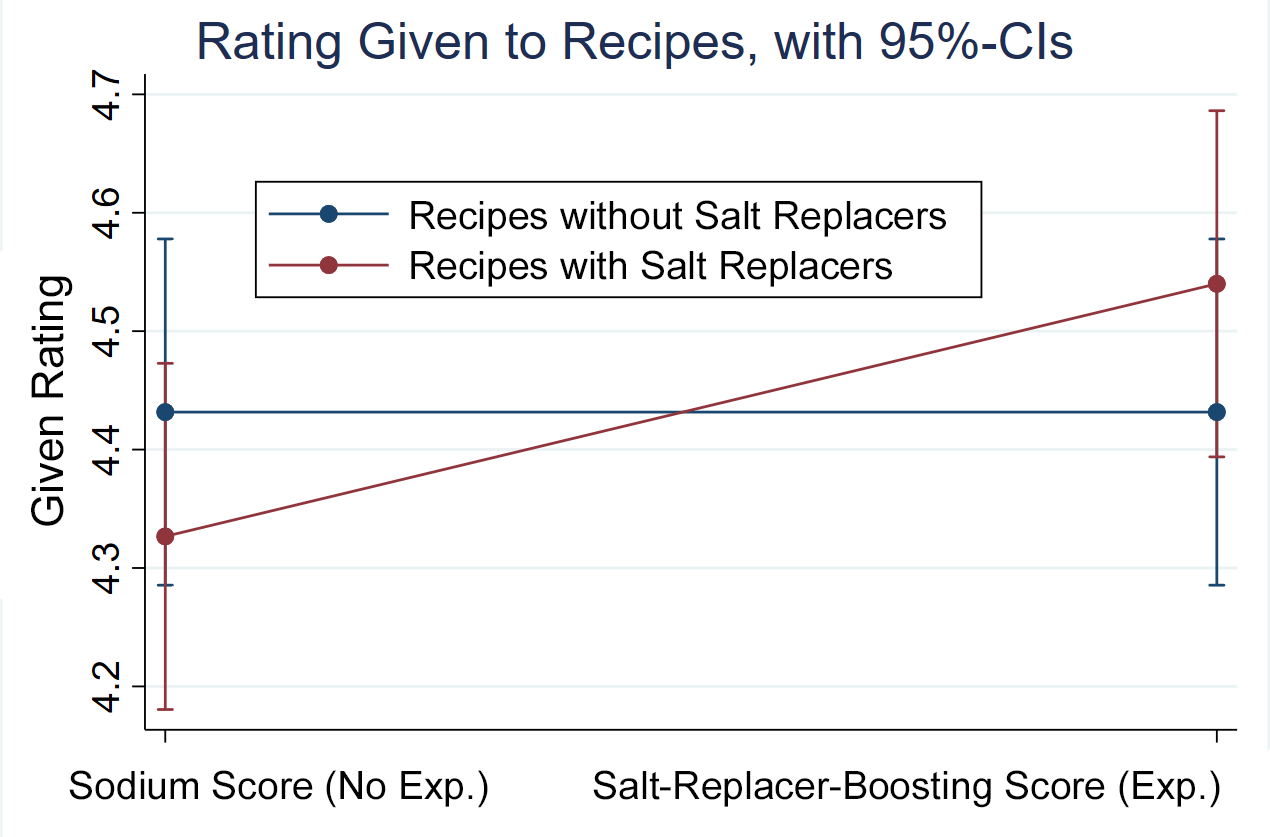}
  \caption{Marginal effects on the ratings for two recipe types with and without salt-replacer ingredients (e.g. oregano) across both conditions (based on sodium score or SR-boosting score). Bars represent 95\%-CIs.}
  \label{figure:condition-effect-salt-replacers}
\end{figure}

It is easier to understand this result by inspecting Figure~\ref{figure:condition-effect-salt-replacers}. While there was no difference across conditions for Non-SR ingredients, the rating given for SR recipes increased around .20 on a scale from 1 to 7. This either suggested that explanations boosted preferences for SR recipes or that they stood out from the larger list of recipes.

\subsection{RQ2: Recipe \& User Characteristics}
We further examined whether other recipe features and user characteristics affected users' food preferences with regard to salty taste. To do so, we predicted the rating given by users to the presented recipes, based on a recipe's taste value, a user's self-reported health with regard to her eating habits, the SR-boosting condition, cooking experience and a user's attitude towards SR ingredients. Table~\ref{tab:multilevel} describes the two random effects models. Model 1 shows that recipes for which a higher taste map value is reported, whether based on sodium score or SR-boosting score, are also more likely to receive a higher rating: $\beta=1.34$, $p<0.001$. This confirmed our expectations that a recipe's salty taste is an important predictor of expected food enjoyment and food preferences.

\begin{table}[]
    \centering
     \caption{Two multilevel regression models predicting the rating given by a user to a presented recipe. Model 1 only examines the relation between the salty taste value and rating, while Model 2 includes more user characteristics and the effects due to the SR-boosting condition. $^{***}p<0.001$, $^{**}p<0.01$, $^{*}~p~<~0.05$.}
    \resizebox{\linewidth}{!}{
        \begin{tabular}{lll}
        \toprule
        \textbf{Factor}    & \textbf{Model 1} & \textbf{Model 2} \\
            & $\beta$ (S.E.) & $\beta$ (S.E.) \\
        \hline
        Presented Taste Map Value & 1.34 (.12)$^{***}$ & 2.64 (.50)$^{***}$ \\
        Self-reported Health &       & .46 (.12)$^{***}$ \\
        \hspace{2mm}Recipe Rating X Self-Health &      & -.37 (.14)$^{**}$ \\
        \\
        Salt-Replacer Attitude & & .19 (.057)$^{**}$ \\
        SR-Boosting Condition   &       & .52 (.34)   \\
        \textit{(Vs Baseline)} & &  \\
        \hspace{2mm}Condition X Self-Health & & -.068 (.074) \\
        \hspace{2mm}Condition X Attitude    & & -.046 (.047) \\
        \\
        Cooking Experience      &           & .050 (.080)   \\
        Age                     &           & -.0045 (.0051)  \\
        \hline
        Constant & 3.72 (.095)$^{***}$ & 1.08 (.49)$^{*}$ \\
        ${\chi}^2$ & 125.04$^{***}$ & 158.44$^{***}$ \\
        ${\rho}$ & 0.25 & 0.23 \\
        $R^{2}$ & 0.038$^{***}$ & 0.072$^{***}$ \\
        \bottomrule
        \end{tabular}
    }
    \label{tab:multilevel}
\end{table}

\begin{figure}[]
    \centering
    \includegraphics[width=1\linewidth]{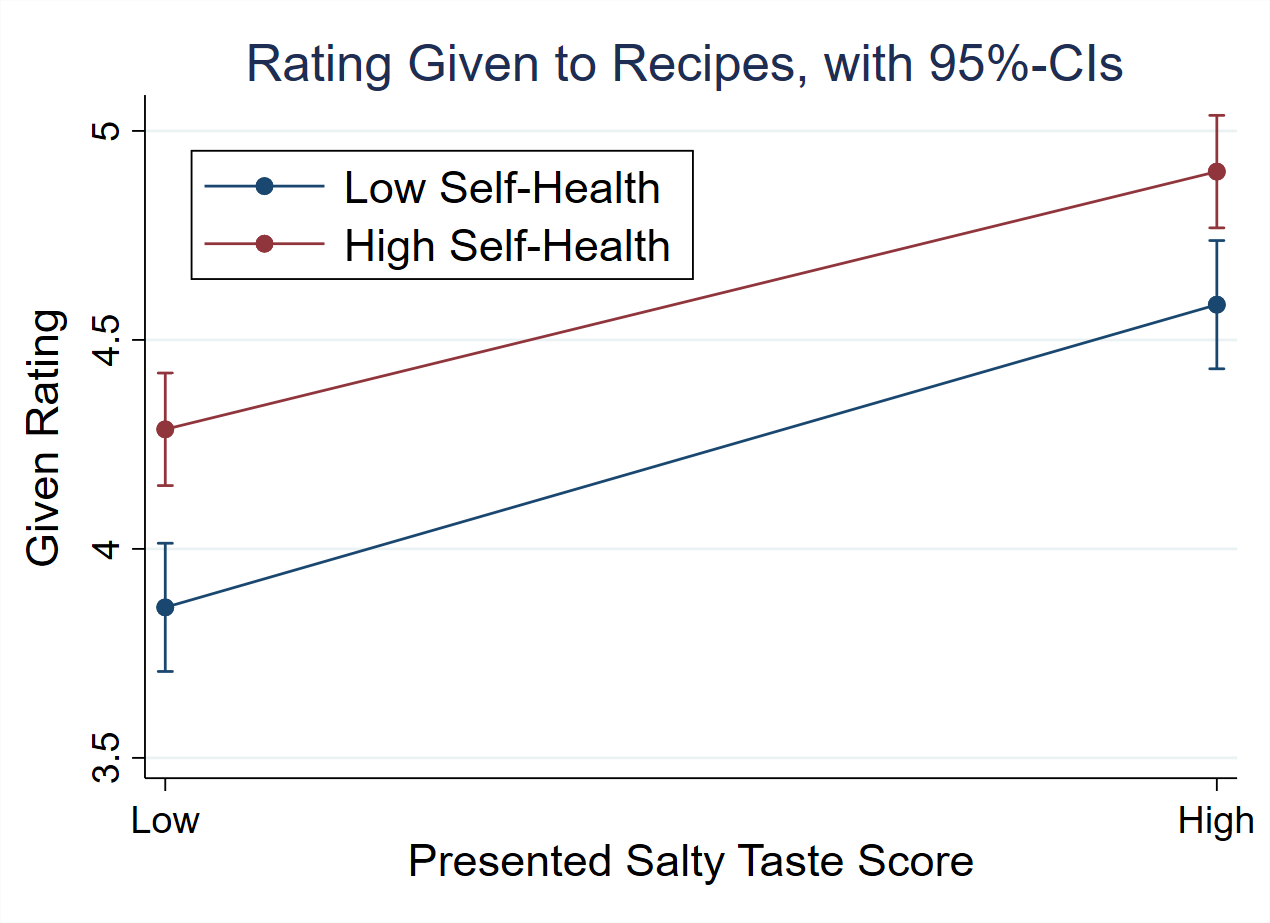}
    \caption{Given rating for presented recipes, based on the presented taste map value and a user's self-reported health. Bars represent 95\%-CIs.}
    \label{fig:taste_score_health}
\end{figure}

Table~\ref{tab:multilevel}, Model 2 expands our baseline model by adding user characteristics, as well as by controlling for our SR-boosting condition. We again observed a positive relation between the presented taste map value and the given rating ($beta=2.68$, $p<0.001$), further confirming our expectations. In addition, Table~\ref{tab:multilevel} shows that users with self-reported health were more likely to give a higher rating: $\beta=.46$, $p<0.01$, which suggested that users who perceived themselves as having a healthy lifestyle still preferred foods that contained higher amounts of salt. However, we also observed an interaction effect between the taste map value and self-reported health, which was understood best by inspecting Figure~\ref{fig:taste_score_health}. While both higher levels of self-reported health and a recipe's taste map value positively affected the rating given, the differences in slopes suggested that the increase was stronger for users with lower self-reported health.

Table~\ref{tab:multilevel} also reports on other user characteristics. We found no relation between a user's cooking experience and the given rating ($p>0.05$), nor did we observe an effect due to age ($p>0.05$). In contrast, we did find that a user's attitude towards SR ingredients positively affected the rating given to recipes ($p<0.01$). Although it seemed sensible that this would mostly apply to recipes that contained such SR ingredients, we observed no interaction effect with the presented taste map value and the SR-boosting condition. Note that analyzing the models in Table~\ref{tab:multilevel} only on recipes that contained SR ingredients led to similar results, except that the interaction effect between taste map value and self-reported health was non-significant.

\section{Summary \& Future Work}\label{sec:conclusions}

For clarity, we summarize our main findings: 
\begin{itemize}
    \item We have examined to what extent food preferences change due to visual and textual explanations on salty taste. We have found that SR recipes can achieve a higher satisfaction of salty food preferences when boosted in the search interface. 
    \item We have also examined to what extent other recipe and user characteristics can affect salty food preferences. Recipes' salty taste appears to be an important predictor of expected food enjoyment, since recipes with a high displayed salty taste value received the highest ratings for fulfilling salty food preferences.
    \item Users with a high level of self-reported health and those with a positive attitude towards SRs provided higher ratings across all recipes, indicating a higher overall expected enjoyment of food. 
    \item We have found evidence that participants with low levels of self-reported health can benefit the most from SR explanations, since they had the largest increase in given ratings. 
\end{itemize}

A number of limitations have to be noted. Due to the novelty of the research domain, our findings have to be interpreted with some caution. Hence, this is one of the first studies on SR ingredients and salty food preferences in a search interface. First of all, there were only three SR ingredients available in our dataset sub-sample, namely garlic, oregano and rosemary. Also, we conducted our study on a very limited number of recipes. A larger study is obviously needed that captures a bigger variety of recipes and SR ingredients.

The study at hand investigated only short-term behavioral change. A longer-term (weeks or months) experiment is needed to justify whether these changes in behavior can be cemented~\cite{Ghawi2014}. 

Furthermore, we did not personalize the search results or provide results based on specific salty food preferences of users. As a first effort, we concentrated on studying the overall effect of boosting SR recipes for the participant population, regardless of initial salty food preferences of individuals. Further research is needed to also understand the effect of personalization~\cite{musto2020towards}.

Future research should look into other axes of the taste map as presented in this paper and especially user goals/dietary restrictions. Sweetness is, for example, another interesting axis that may be worthwhile to explore further, as high sugar intake also increases health risks. Finally, more research is required to further examine how to boost healthier recipes in search interfaces by means of novel interface interventions, without sacrificing user satisfaction~\cite{Trattner2017a}.

\begin{acknowledgments}
  This research is in parts funded by MediaFutures partners and the Research Council of Norway (grant numbers 309339 and 310468).
\end{acknowledgments}

\bibliography{bibliography}

\appendix

\section{Online Resources}

The prototype source code and Stata dataset for the online experiment is available at \url{https://git.app.uib.no/foodrecsys/salty-food-preferences}. The prototype code is an adaptation of Martijn Willemsen’s MouseLabWeb.

\end{document}